\documentclass[12pt]{article}
\usepackage[utf8]{inputenc}
\usepackage{graphicx}
\usepackage{color}
\usepackage{geometry}  
\usepackage{parskip}   
\usepackage{amsmath, amssymb} 
\usepackage{setspace} 
\usepackage{booktabs} 
\usepackage[round]{natbib} 
\DeclareMathOperator{\E}{\mathbb{E}}
\usepackage{float}
\usepackage[misc]{ifsym}
\usepackage{tabularx}
\usepackage{subcaption}
\usepackage{siunitx}
\usepackage[most]{tcolorbox}
\usepackage{eurosym}

\newcommand{\bs}[1]{\ensuremath{\boldsymbol{#1}}} 

\newcolumntype{Y}{>{\centering\arraybackslash}X}

\begin{document}

\title{A tool to determine the degrees of freedom in tree-structured varying coefficient models} 
\author{Nikolai Spuck\footnote{Core Facility Biostatistics, Central Institute of Mental Health, Medical Faculty Mannheim, Heidelberg University, Mannheim, Germany.} { }and Moritz Berger$^{*}$}
\date{\today}

\maketitle
\vspace{-0.5cm}
\begin{abstract}
\noindent The tree-structured varying coefficient (TSVC) model is a flexible approach for generalized regression, where the linear effects of the covariates are allowed to vary with the values of effect modifiers. Relevant effect modifiers and interactions are identified using recursive partitioning. In TSVC models, analogously to other semi- and nonparametric regression approaches, one needs to account for the cost of data-driven model building when deriving the model degrees of freedom (DoF).
To address this issue, we develop an easy-to-apply formula to approximate the DoF of a TSVC model. This formula is employed for model selection based on the Bayesian information criterion (BIC) and compared to the naive solution, setting the DoF to the number of free model parameters, in a simulation study. To illustrate the proposed DoF method, TSVC models using BIC-based selection were fitted to data from the Survey of Health, Ageing, and Retirement in Europe. Results indicated that calculation of the DoF using the proposed formula resulted in more accurate selection results with improved predictive ability.       
\end{abstract}

\vspace{1ex}
\hrule width0.4\textwidth
\hspace{0.5ex}\footnotesize{\Letter\ \ Nikolai Spuck} 
\vspace{-0.5ex}

\vspace{-2.5ex}
\hspace{4.1ex}\footnotesize{\ Nikolai.Spuck@zi-mannheim.de}
\vspace{-1ex}
\normalsize

\section{Introduction}

In statistics, the model \textit{degrees of freedom} (DoF) quantify the flexibility of a fitting procedure, which is an essential prerequisite in many practical applications, in particular, with regard to model selection. For instance, information criteria such as the Akaike information criterion (AIC; \citealp{Akaike1974}), the Bayesian information criterion (BIC; \citealp{Schwarz1978}) and the deviance information criterion (DIC; \citealp{Spiegelhalter2002}), which are designed to select an optimal model from a sequence of nested models with increasing complexity, employ the model DoF in order to solve the bias-variance trade-off. 

In general, the model DoF are defined as the \textit{effective number of parameters} in a statistical model \citep{Efron1986, Hastie1990}. In a classical linear regression model, this simply equals the absolute number of free parameters (i.e. the intercept and the covariate effects). For more complex modeling approaches like semi- and nonparametric regression or regression models involving variable selection methods, how to determine the model DoF is not immediately obvious. 
Let us follow the definition by \citet{Efron1986} and assume uncorrelated, normally-distributed observations $\bs{y}=(y_1,...,y_n)^\top$ of the form
\begin{equation*}
\bs{y} = \bs{\mu} + \bs{\varepsilon} \quad \text{with} \; \bs{\varepsilon}\sim MVN(\bs{0}_n, \sigma^2I_{n\times n})\, ,
\end{equation*}
where $\bs{\mu}\in \mathbb{R}^{n}$ denotes the expectation vector, $\sigma^2 \in \mathbb{R}^+$ denotes the variance of the error terms $\bs{\varepsilon}$, $\bs{0}_n$ is the $n$-dimensional zero vector, and $I_{n\times n}$ is the $n\times n$ identity matrix.
For any method $M:\mathbb{R}^n \to \mathbb{R}^n$ that generates fitted values $\hat{\bs{\mu}}(\bs{y}) = M(\bs{y})$ the DoF are given by
\begin{equation}
\label{gdf}
df^{M}(\bs{\mu}) = \frac{1}{\sigma^2}\sum_{i=1}^{n} \text{Cov}(\hat{\mu}_i(\bs{y}), y_i)\, .
\end{equation}
The definition in Equation~\eqref{gdf} is also referred to as the generalized DoF and has been studied extensively by \citet{Ye1998} and \citet{Efron2004}. \citet{Ye1998} investigated the DoF for several complex modeling procedures using a Monte Carlo (MC) approach, and \citet{Efron2004} explored the relationship between cross-validation and information criteria, which penalize higher DoF. 

As mentioned above, complex semi- or nonparametric models may often exhibit higher DoF that exceed the number of free model parameters. This is because of the increased flexibility induced by the data-driven model building, which corresponds to a search for the optimal model through the space of eligible model specifications. \citet{Tibshirani2015} defined the cost of this search as the \textit{search DoF}. He postulated that the search DoF are equal to the difference between a model's DoF and its number of free parameters. In the last years, the model DoF were also the subject of research by several other authors. Among others, \citet{Zhou2007} and \citet{Tibshirani2012} showed that the DoF in regularized regression models with LASSO penalty are equal to the number of free non-zero parameters. \citet{Hansen2014} considered the DoF for nonlinear least squares estimation using a geometric approach. In addition, a theorem to determine the DoF for least squares regression with linear constraints or quadratic penalties was derived by \citet{Chen2020}, and \citet{Mentch2020} investigated the DoF in random forests. Just recently, \citet{Wang2024} further explored the concept of search cost in adaptive models and investigated the DoF in spline- and tree-based modeling approaches using MC simulation. Despite these advances, closed-form expressions for calculating the DoF are currently unavailable for many data-driven modeling procedures.

The focus of this work is on estimating the DoF in tree-based regression models. More specifically, the tool developed here is tailored to on an extension of varying coefficient models using tree structures \citep{Berger2019}. Tree-structured varying-coefficient (TSVC) models are embedded  into a generalized regression framework supporting outcomes on continuous and discrete scales, and enable flexible modeling of interaction effects in a data-driven way. Specifically, TSVC models allow the linear effect of each covariate to change with the value of one or several other variables, the so called \textit{effect modifier(s)}. For this, the TSVC fitting procedure applies recursive partitioning to identify relevant effect modifiers, which yields a separate tree for the effect of each covariate with a partition-specific coefficient in each leaf node. To determine the size of the trees, \citet{Berger2019} originally proposed to apply permutation tests as a stopping criterion in the tree building algorithm. More recent works suggested to use the cross-validated log-likelihood to optimize the number of splits or to determine an optimal minimal node size \citep{Spuck2023, Spuck2026}. As a third alternative, \citet{Berger2018} and \citet{Spuck2025} applied the BIC to select an optimal model from the sequence of nested models fitted within the TSVC framework. While permutation tests and cross-validation may lead to substantial computational costs, the BIC offers a convenient solution but does not adequately account for the search cost induced by the tree building when setting the DoF equal to the number of free parameters. In TSVC models, this approximation turned out to be sufficient only in lower dimensional settings (with a small number of effect modifiers), which was the main driver for the present work. 



Here, we investigate this issue in a general manner and relate the DoF of a TSVC model to the number of splits, the number of covariates, and the sample size of the data used for fitting. We provide an easy-to-apply formula, which can be used to approximate the DoF in any data setting. For this purpose, we first calculated the generalized DoF based on Equation~\eqref{gdf} using a MC approach and fitted a flexible \textit{multivariable fractional polynomial} (MFP; \citealp{Royston1994}) regression model to the resulting data in a second step (detailed in Section~3).   

The remainder of this article is structured as follows: In Section 2, the concept of TSVC models and the fitting procedure are outlined. In Section 3, the formula for approximating the DoF in TSVC models is derived. This formula is applied for model selection via BIC and compared to the naive solution, which neglects the search cost of the model building, in a simulation study (Section 4). An application to data from the Survey of Health, Ageing and Retirement in Europe (SHARE) is presented in Section 5. Results are summarized and discussed in Section 6.

\section{Tree-structured varying coefficient models}

Let $ Y$ be the outcome variable of interest and $\bs{X} = (X_1,\dots , X_p)^\top $ denote the vector of covariates. We assume that $X_j$ are ordinally or metrically scaled, or dummy-coded representations of nominal variables. In generalized regression models, the conditional distribution of the outcome $Y$ given covariates $\bs{X}$ is assumed to belong to the exponential family. The expected outcome is then related to the covariate vector in the form $\E (Y|\, \bs{X}) = g^{-1}(\eta (X))$, where $g(\cdot)$ denotes a suitable link function and $\eta(\cdot)$ denotes the predictor function. In the simplest case, the predictor function is a linear combination of the covariates $\eta(\bs{X})=\bs{X}^\top\bs{\beta}$, with the vector of regression coefficients $\bs{\beta}=(\beta_1,\hdots,\beta_p)^\top$. The more flexible predictor function of the TSVC model introduced by \citet{Berger2019} has the general form 
\begin{equation}
\eta (\bs{X}) = \beta_0 + \beta_1 (\bs{X}_{[-1]}) X_1 + \dots + \beta_p (\bs{X}_{[-p]})X_p\, ,
\end{equation}
where $\bs{X}_{[-j]}=\{X_{1} \dots X_{j-1},X_{j+1},\dots, X_p\}$ denotes the set of all covariates excluding $X_j$. By design, the effect of each covariate can be modified by each other covariate except itself. In TSVC, each of the functions $\beta_j(\cdot )$ comprises varying coefficients of $X_j$ that are determined by a tree structure. That is, each function $\beta_j(\cdot )$ sequentially partitions the observations into disjoint subsets $N_{jm},\, m=1,\hdots,M_j$, referred to the predictor space of $\bs{X}_{[-j]}$ (the potential effect modifiers), and assigns a separate node-specific coefficient for $X_j$ to each partition $N_{jm}$. These functions can be written as 
\begin{equation}
\beta_{j}(\bs{X}_{[-j]}) = \sum_{m}^{M_{j}} \beta_{jm}\, I(\bs{X}_{[-j]}\in N_{jm})\, ,
\end{equation}
where $I(\cdot )$ is the indicator function. Each coefficient is derived from a set of binary splitting rules, which successively partition a parent node into two child nodes (for the basic concepts on recursive partioning, see also \citealp{Hastie2009}). The starting point is a model with non-varying linear effects, only. Then, as an example, the first split yields a model with predictor
\begin{align*}
\eta^{[1]} (\bs{X}) = &\, \beta_0^{[1]} + \beta_{1}^{[1]}X_{1} + \dots +\left(\beta_{j1}^{[1]}I(X_{k}\leq c_k) + \beta_{j2}^{[1]}I(X_{k}> c_k)\right)X_{j} \nonumber \\
& + \dots + \beta_p^{[1]} X_{p}\, ,
\end{align*}
where $c_k$ is the split point in effect modifier $X_k$ selected by the algorithm regarding the effect of $X_j$. The parameter $\beta_{j1}^{[1]}$ is the linear effect of $X_j$ in partition $\{X_{k} \leq c_k\}$ and $\beta_{j2}^{[1]}$ is the linear effect of $X_{j}$ in partition $\{X_{k}>c_k\}$.  Hence, after the first step, the varying coefficient of $X_j$ is determined by $\beta_j^{[1]}(X_{k}) = \beta_{j1}^{[1]}I(X_{k}\leq c_k) + \beta_{j2}^{[1]}I(X_{k}> c_k)$. In the next step, either a different coefficient is selected for splitting or the same coefficient is further modified. If the coefficient of variable $X_\ell$ is split in $X_{r}$ at split point $c_r$ this yields the predictor
\begin{align*}
\eta^{[2]} (\bs{X} ) =&\, \beta_0^{[2]} + \beta_{1}^{[2]}X_{1} + \dots +\left(\beta_{j1}^{[2]}I(X_{k}\leq c_k) + \beta_{j2}^{[2]}I(X_{k}> c_k)\right)X_j   \nonumber \\
 & +  \left(\beta_{\ell1}^{[2]}I(X_{r}\leq c_r) + \beta_{\ell2}^{[2]}I(X_{r}>c_r)\right)X_{\ell} +\dots + \beta_p^{[2]} X_{p}\, ,
\end{align*} 
where $\beta_{\ell1}^{[2]}$ denotes the effect of $X_\ell$ in $\{X_{r}\leq c_r\}$ and $\beta_{\ell2}^{[2]}$ denotes the effect of $X_{\ell}$ in $\{X_{r}>c_r\}$. That is, the varying coefficient of $X_\ell$ has the form $\beta_{\ell}^{[2]}(X_{r}) = \beta_{\ell1}^{[2]}I(X_{r}\leq c_r) + \beta_{\ell2}^{[2]}I(X_{r}>c_r)$. Further splits are performed analogously until a predefined stopping criterion is met (see below for details). In each step, a so far non-varying linear effect turns into a varying coefficient or an already selected varying coefficient is split further. 

\subsection*{Fitting procedure}

In each step of the tree building algorithm, the best splitting rule from among all possible combinations of covariate $X_j$, respective candidate effect modifier $X_k, k \neq j$, and split point is selected, starting from a linear predictor without varying coefficients. For this, all candidate models with one additional split are evaluated and the best-performing one that yields the smallest deviance is selected. In generalized regression models the deviance is a quite natural measure of the model fit. This criterion is also equivalent to minimizing the entropy, which has already been used as a splitting criterion in the early days of tree construction \citep{Breiman1984}. Note that, in contrast to common trees, in each step of the algorithm all the observations are used to derive new estimates of the model parameters. This ensures that one obtains valid estimates of the different model components together with the splitting rule.

To determine the optimal number of splits and hence the size of the trees, \citet{Berger2019} proposed an early stopping strategy based on permutation tests. This offers an approximate solution to control the global type I error rate (that is, the proportion of falsely identified covariates with varying coefficients). 
An attractive alternative, is post-pruning, where a large model is grown first and is then pruned to an adequate size to avoid overfitting. Running the stepwise TSVC algorithm (with a sufficiently large number of splits) yields a sequence of nested models that can be assessed with regard to their goodness of fit using a likelihood-based criterion. Earlier works considered the cross-validated log-likelihood as selection criterion for post-pruning~\citep{Spuck2023, Spuck2026}. Here, we focus on the BIC, as it is easy to use and widely applied for model selection by practitioners. The BIC of a TSVC model $M^{[s]}$ is given by
\begin{equation}
BIC\left (M^{[s]} \right ) = - 2\ln \big( L^{M^{[s]}} \big) +  \ln (n) \, df^{M^{[s]}}(\bs{\mu})\, ,
\end{equation}
where $L^{M^{[s]}}$ denotes the maximized value of the likelihood function of the model with $s$ splits and $df^{M^{[s]}}$ are the corresponding model DoF. A naive solution is to set the the model DoF equal to the sum of the number of covariates $p$ and the number of performed splits $s=\sum_{j=1}^{p}(M_j-1)$, which corresponds to the number of free parameters. However, this does not account for the search cost induced by the tree-building algorithm. In the next section, we derive a formula to estimate the DoF, which is tailored to TSVC models and builds on the simulation-based approach by \citet{Wang2024}. 

\section{Estimating the degrees of freedom}

To derive an estimate of the DoF in a TSVC model, we perform a two-step approach. In the first step, we approximate the theoretical DoF defined in Equation~\eqref{gdf} using MC simulations as described in \citet{Wang2024}. We consider several combinations of potentially influential factors: number of splits (\textit{s}), number of covariates (\textit{p}), and sample size of the data used for fitting (\textit{n}). In the second step, we fit a regression model, where the estimated DoF serve as the response variable and the aforementioned factors are included as explanatory variables. From this we build a regression formula that can directly be applied as a tool to calculate the DoF given these factors. 

\subsection{Monte Carlo simulation}

Following \citet{Wang2024}, we resort to a MC approach to approximate the theoretical DoF. Specifically, we perform the following algorithm:

\begin{enumerate}
    \item \textbf{Determine the expectation vector:} If the design matrix $\bs{X}\in \mathbb{R}^{n\times p}$ and the true coefficient vector $\bs{\beta}\in \mathbb{R}^p$ are known, set $\bs{\mu} = \bs{X}\bs{\beta}$. Otherwise, set $\bs{\mu} = \bs{0}_n$. 
    \item \textbf{Data generation:} Simulate $m$ times $n$ independent observations $y_{ij} \sim N(\mu_i, 1),\; i = 1,\dots , n,\; j=1,\dots, m$.
    \item \textbf{Model fitting:} Fit $m$ TSVC models to each of the data sets $\{\bs{y}_{j}, \bs{X}\},\; j=1,\dots, m$, and generate the corresponding fitted values $\hat{\mu}_i^{(j)}$.
    \item \textbf{Calculate the empirical DoF:} Let $\hat{\bs{\mu}}_i^\top=(\hat{\mu}_i^{(1)},\hdots, \hat{\mu}_i^{(m)})$. Calculate and return the sum of the sample covariances following Equation~\eqref{gdf}:
    $$ \widehat{df}^{M^{[s]}}(\bs{\mu}) = \sum_{i=1}^{n}\widehat{\text{Cov}}(\bs{\hat{\mu}}_i, \bs{y}_i)\, .$$
\end{enumerate}

We applied the algorithm above to calculate the DoF in 20 different data settings. We simulated data with $p\in \{2, 4, 6, 8, 10\}$ standard normally distributed covariates of sample sizes $n\in\{100, 400, 700, 1000\}$. Note that, analogously to other data-driven modeling approaches, a TSVC model's DoF also depend on the specific form of the data generating process (DGP). However, \citet{Ye1998} found only minor DGP-based differences in the DoF for other tree-based methods and argued that a DGP without covariate effects yields the highest DoF, that is, results in the most parsimonious model. In all of our considered data settings the expectation of the response variable was, therefore, set to $\bs{\mu} = \bs{0}_n$. This issue with regard to model parsimony is discussed in more detail in Section 6.

For all data settings, a TSVC model with a maximal number of $S_{\text{max}}=5$ splits was fitted and the DoF were calculated as an average from $R=10$ empirical DoF resulting from the outlined algorithm based on $m = 100$ MC replications. Importantly, the DoF were considered for all nested TSVC models with $s = 1,\dots, 5$ splits. Overall, this simulation setup resulted in empirical DoF values for 100 different combinations of $s$, $p$ and $n$. In the following, we will refer to these results as the \textit{MC approximated DoF}.

\begin{table}[ht]
\centering
\caption{MC approximated DoF of a TSVC model. Estimates calculated as an average of the results from $R=10$ applications of the MC algorithm with $m=100$ MC replications each for the number of variables $p\in\{2, 4, 6, 8, 10\}$, sample size $n\in \{100, 400, 700, 1000\}$ and number of splits $s\in \{1, \dots, 5\}$ with standard errors in brackets. }
\label{edf}
{\footnotesize
\begin{tabularx}{\textwidth}{Yccccccc}
\hline
&&$s$ & 1 & 2 & 3 & 4 & 5 \\
$p$ & $n$ & \\
\hline
2 & \ 100  & & \ 7.41 (0.14) & 11.26 (0.14) & 14.26 (0.15) & 16.90 (0.17) & 19.27 (0.19) \\
 & \ 400  & & \ 7.34 (0.11) & 11.40 (0.09) & 14.48 (0.11) & 17.23 (0.11) & 19.62 (0.11) \\
 & \ 700  & & \ 7.55 (0.11) & 11.56 (0.16) & 14.67 (0.16) & 17.44 (0.17) & 19.93 (0.19) \\
 & 1000 & & \ 7.40 (0.13) & 11.46 (0.13) & 14.58 (0.15) & 17.38 (0.17) & 19.82 (0.18) \\
\hline
4 & \ 100  & & 12.45 (0.16) & 18.32 (0.17) & 23.48 (0.19) & 28.08 (0.20) & 32.25 (0.23) \\
 & \ 400  & & 12.79 (0.16) & 19.39 (0.20) & 25.36 (0.19) & 30.91 (0.22) & 36.19 (0.24) \\
 & \ 700  & & 12.73 (0.10) & 19.54 (0.09) & 25.59 (0.14) & 31.36 (0.21) & 36.87 (0.24) \\
 & 1000 & & 12.85 (0.16) & 19.55 (0.19) & 25.77 (0.24) & 31.63 (0.26) & 37.33 (0.25) \\
\hline
6 &\  100  & & 15.89 (0.17) & 22.96 (0.17) & 28.99 (0.19) & 34.42 (0.20) & 39.28 (0.22) \\
 & \ 400  & & 16.46 (0.20) & 24.39 (0.24) & 31.66 (0.27) & 38.55 (0.30) & 45.03 (0.28) \\
 & \ 700  & & 16.50 (0.15) & 24.62 (0.19) & 32.21 (0.22) & 39.35 (0.25) & 46.16 (0.23) \\
 & 1000 & & 16.74 (0.12) & 24.90 (0.16) & 32.55 (0.20) & 39.64 (0.25) & 46.64 (0.28) \\
\hline
8 & \ 100  & & 18.99 (0.13) & 26.96 (0.13) & 33.78 (0.14) & 39.77 (0.16) & 45.16 (0.19) \\
 & \ 400  & & 19.64 (0.18) & 28.71 (0.24) & 36.94 (0.28) & 44.69 (0.31) & 52.09 (0.35) \\
 & \ 700  & & 19.98 (0.14) & 29.18 (0.18) & 37.75 (0.21) & 45.77 (0.23) & 53.53 (0.27) \\
 & 1000 & & 19.94 (0.15) & 29.25 (0.15) & 37.89 (0.17) & 46.04 (0.19) & 53.90 (0.21) \\
\hline
10 & \ 100  & & 21.84 (0.12) & 30.48 (0.13) & 37.89 (0.13) & 44.26 (0.16) & 49.90 (0.17) \\
 & \ 400  & & 22.66 (0.23) & 32.58 (0.26) & 41.63 (0.30) & 50.07 (0.33) & 58.15 (0.34) \\
 & \ 700  & & 22.90 (0.14) & 32.99 (0.17) & 42.29 (0.21) & 51.09 (0.24) & 59.58 (0.25) \\
 & 1000 & & 22.92 (0.15) & 33.09 (0.15) & 42.48 (0.19) & 51.45 (0.22) & 60.01 (0.24) \\
\hline
\end{tabularx}
}
\end{table}

From the results in Table~\ref{edf} it is immediately apparent that the MC approximated DoF are much larger than the number of free parameters of the TSVC model. The estimated search cost or search DoF (MC approximated DoF minus the number of free model parameters) heavily increase with the number of performed splits~$s$ and the number of covariates~$p$, and also with the sample size $n$. For instance, in the case of $n=100$, the search cost of a TSVC model with $p=2$ covariates and $s=1$ splits are approximated by $7.41-4 = 3.41$, whereas the search cost for a TSVC model with $p=10$ and $s=5$ is approximately $49.9-16=33.9$.  The effect of the sample size on the search cost appeared to be negligible if the number of splits and the number of covariates were low but increased with both of these values. Specifically, for a TSVC model with $p=2$ and $s=1$ the search cost are estimated as $3.41$ if $n=100$ and $3.40$ if $n=1000$. For a TSVC model with $p=10$ and $s=5$, however, a much larger difference in the search cost between the sample sizes $n=100$ (search cost: $33.90$) and $n=1000$ (search cost: $44.01$) was observed. 

\subsection{Regression formula}

The framework of multivariable fractional polynomial (MFP) regression originally proposed by \citet{Royston1994} and further refined by \citet{Sauerbrei1999} enables variable selection combined with the selection of a suitable functional form for continuous covariates in multivariable regression analysis. A detailed description of the MFP method is outlined in the Appendix. 

Here, we used this procedure and fitted an MFP model with degree $d\leq 2$ to the data from Table~\ref{edf} in order to derive an easy-to-apply formula for directly calculating the DoF of a TSVC model. In this model, the MC approximated DoF served as outcome and the number of splits $s$, number of covariates $p$, and sample size $n$ as explanatory variables. When fitting the MFP model, we allowed for all second- and third-order interactions between these potentially influential factors. The error levels for variable selection and testing the functional forms were set to $\alpha = 0.05$.

\begin{tcolorbox}[enhanced, colframe=black, colback=white]
The predictor of the fitted MFP model for calculating the DoF of a TSVC model $M^{[s]}$ has the form
\begin{equation}
\label{mfpdf}
\hat{df}_{\text{MFP}}^{M^{[s]}}\left(\bs{\mu}\right) = 2.13 +2.02\, s+ 1.26\, p+0.61\, p\,s + 0.16\cdot 10^{-3}\, p \, s \, n\,, 
\end{equation}
with number of splits $s\geq 1$, number of covariates $p\geq 2$, and sample size $n$ of the data the model was fitted to.
\end{tcolorbox}

\begin{figure}[!t]
\centering
\includegraphics[width = 0.55\textwidth]{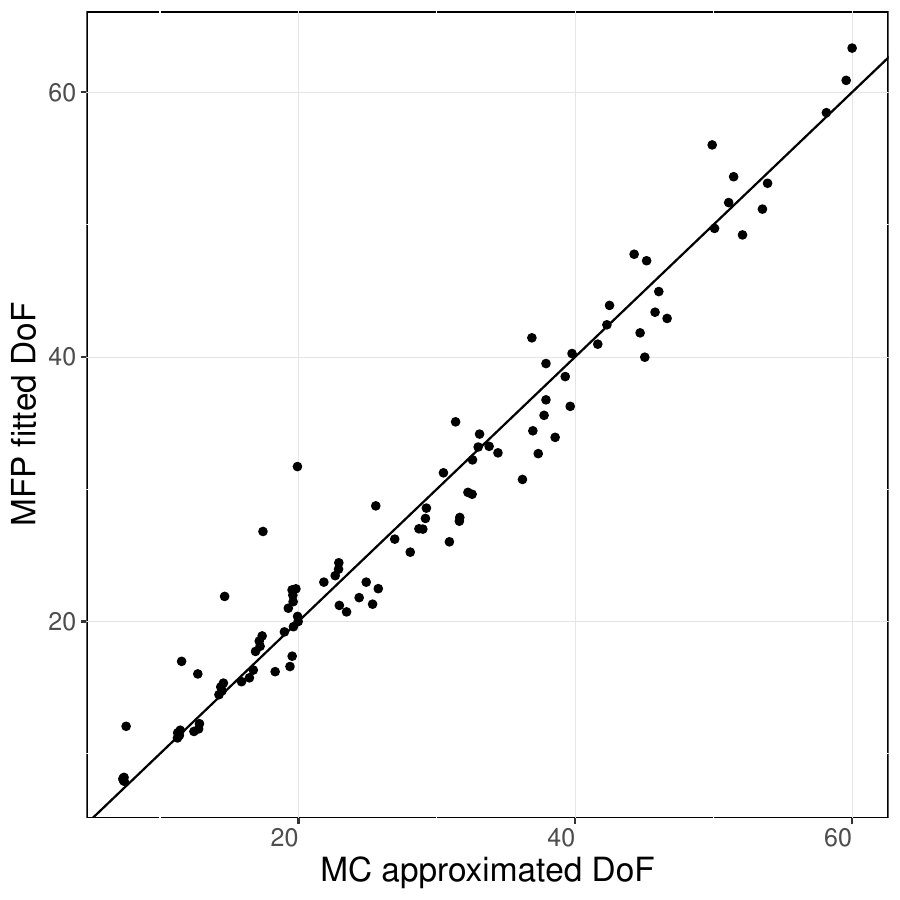}
\caption{Results of the MPF model. The figure shows the approximated DoF based on the MC algorithm on the x-axis plotted against the fitted DoF from the MFP model given in Equation~\eqref{mfpdf} on the y-axis.}
\label{fittedDoF}
\end{figure}
Equation~\eqref{mfpdf} shows a strong association between the DoF and~$s$ as well as~$p$.
The MPF model selected linear effects for $s$ and $p$, whereas no main effect of $n$ on the DoF was detected. The algorithm also found a second-order interaction between $s$ and $p$ with a linear positive effect, which indicates a disproportionate increase in DoF for large $s$ and $p$. Further, the third-order interaction between $s$, $p$, and $n$ was selected, revealing the effect of the sample size with large number of splits and covariates. Figure~\ref{fittedDoF} shows that the fitted DoF from the MFP model align closely with the approximated values from the MC algorithm that were used for model fitting. Overall, the model explained $R^2=95.5\%$ of the variance in the MC approximated DoF.

\section{Simulation study}

To assess the practicability of the proposed regression formula in Equation~\eqref{mfpdf} for the DoF of a TSVC model, we considered several scenarios in a simulation study. In particular, the objective of the simulation study was (i) to investigate the performed number of splits and the predictive performance of a TSVC model with pruning by BIC, and (ii) to compare the resulting models to those selected based on the naive definition of the DoF neglecting the search cost.    

We considered three simulation scenarios with a different number of covariates. \textit{Scenarios 1}, \textit{2}, and \textit{3} consider TSVC models with $p=2$, $p=6$, and $p=10$ standard normally distributed covariates, respectively. For each scenario DGPs with the true number of splits $s_{\text{DGP}}\in\{0, 1, 2, 3\}$ in the linear effects of the covariates and sample sizes $n\in \{100, 400, 1000\}$ were considered, resulting in $3\times 3=9$ settings per scenario. Furthermore, in a fourth scenario based on our application to the SHARE data, additional settings with a varying number of splits $s_{\text{DGP}}\in \{0, 2, 4, 6\}$ with sample size $n=2985$ and number of covariates $p=4$ were considered. In each setting, 100 replications were performed.

The TSVC models were fitted with post-pruning based on the BIC. The following approaches were used to determine the DoF:
\begin{enumerate}
\item[(i)] the naive approach, taking only the number of free parameters $p+s+1$ into account (\textit{Naive}),
\item[(ii)] applying the proposed formula in Equation~\eqref{mfpdf}, derived from MFP regression on the MC approximated DoF (\textit{MC MFP}),
\item[(iii)] the MC approximated DoF from Table~\ref{edf} based on a DGP with only null effects (\textit{MC Null}), and
\item[(iv)] the MC approximated DoF based on the true DGP (\textit{MC DGP}). Note that these are not available in practice, where the true DGP is not known, and serves a reference only.
\end{enumerate}

All TSVC models were fitted with a maximal number of $S_{\text{max}}=5$ splits in the first three scenarios, and $S_{\text{max}}=10$ in the application-based fourth scenario, before post-pruning was applied. 

For evaluation, the average number of splits performed by the TSVC algorithm was calculated. In addition, the performance of the models was evaluated based on the predictive log-likelihood calculated on a test sample of the same size as the sample used for fitting. 

\subsection{Simulation with two covariates}

In scenario 1, a two-covariate DGP of the form 
\begin{equation*}
Y = \beta_1(X_{2}) \,X_{1} + \beta_{2}X_{2} +\varepsilon
\end{equation*}
was considered, where the error term $\varepsilon$ followed a standard normal distribution. Accordingly, the linear effect of $X_1$ varied with $X_2$ and was determined by a tree structure. The effect of $X_1$ was set to $\beta_{1}(X_{2}) =0$ for $s_{\text{DGP}} = 0$, $\beta_{1}(X_{2}) =I(X_2>0)$ for $s_{\text{DGP}} = 1$, $\beta_{1}(X_{2}) =I(X_2>0) + 2\,I(X_2>0.675)$ for $s_{\text{DGP}} = 2$, and $\beta_{1}(X_{2}) =I(X_2>0) + 2\,I(X_2>0.675) - I(X_2\leq 0.675)$ for $s_{\text{DGP}} = 3$. The linear main effect of $X_2$ was set to $\beta_2 = 0$. 

\begin{table}[ht]
\centering
\caption{Results of the simulation study: Number of splits (scenario 1). Average number of splits (with standard deviations in brackets) of TSVC models including $p=2$ covariates with different methods for determining the DoF (Naive, MC MFP, MC Null, and MC DGP) for the BIC-based post-pruning fitted to samples of size $n\in\{100, 400, 1000\}$, where the true number of splits in the DGP was $s_{\text{DGP}}\in \{0, \dots, 3\}$.}
\label{splits_p2}
\begin{tabularx}{\textwidth}{llYYYY}
\hline
$n$ & $s_{\text{DGP}}$ & \multicolumn{4}{c}{DoF approach}\\ 
& & Naive & MC MFP & MC Null & MC DGP \\
\hline
100  & 0 & 1.03 (1.37) & 0.00 (0.00) & 0.00 (0.00) & 0.00 (0.00) \\
     & 1 & 1.92 (1.13) & 0.57 (0.52) & 0.65 (0.48) & 0.84 (0.37) \\
     & 2 & 2.99 (1.06) & 1.49 (0.52) & 1.36 (0.50) & 1.76 (0.43) \\
     & 3 & 3.54 (0.93) & 2.04 (0.42) & 2.04 (0.53) & 2.76 (0.47) \\
\hline
400  & 0 & 0.41 (0.87) & 0.00 (0.00) & 0.00 (0.00) & 0.00 (0.00) \\
     & 1 & 1.40 (0.68) & 1.00 (0.00) & 1.00 (0.00) & 1.00 (0.00) \\
     & 2 & 2.72 (0.81) & 2.06 (0.24) & 2.08 (0.31) & 2.05 (0.22) \\
     & 3 & 3.84 (0.77) & 2.96 (0.45) & 3.08 (0.54) & 3.11 (0.31) \\
\hline
1000 & 0 & 0.26 (0.65) & 0.00 (0.00) & 0.00 (0.00) & 0.00 (0.00) \\
     & 1 & 1.36 (0.64) & 1.00 (0.00) & 1.00 (0.00) & 1.00 (0.00) \\
     & 2 & 2.75 (0.76) & 2.21 (0.41) & 2.24 (0.43) & 2.28 (0.45) \\
     & 3 & 3.99 (0.70) & 3.33 (0.49) & 3.48 (0.54) & 3.65 (0.52) \\
\hline
\end{tabularx}
\end{table}

Table \ref{splits_p2} shows that using the naive DoF led to an inflated number of splits, in particular in settings with small sample size and a low number of true splits ($n = 100$ and $s_{\text{DGP}} \leq 2$). Conversely, with the MC-based approaches (MC MFP, MC Null, and MC DGP) the TSVC algorithm approximated the true number of splits quite well, with the tendency to underestimate the number of splits for $n=100$. For  the DGP with $s_{\text{DGP}} = 2$ or $s_{\text{DGP}} = 3$ splits in cases of larger sample size ($n=400$ and $n=1000$), the number of splits performed with the MC-based approaches even slightly exceeded the true number. Overall, the proposed MC MFP approach yielded highly reasonale results (parsimonious models in the settings without splits and good approximation in the varying-coefficients settings). 

\begin{figure}[!t]
\centering
\includegraphics[width = \textwidth]{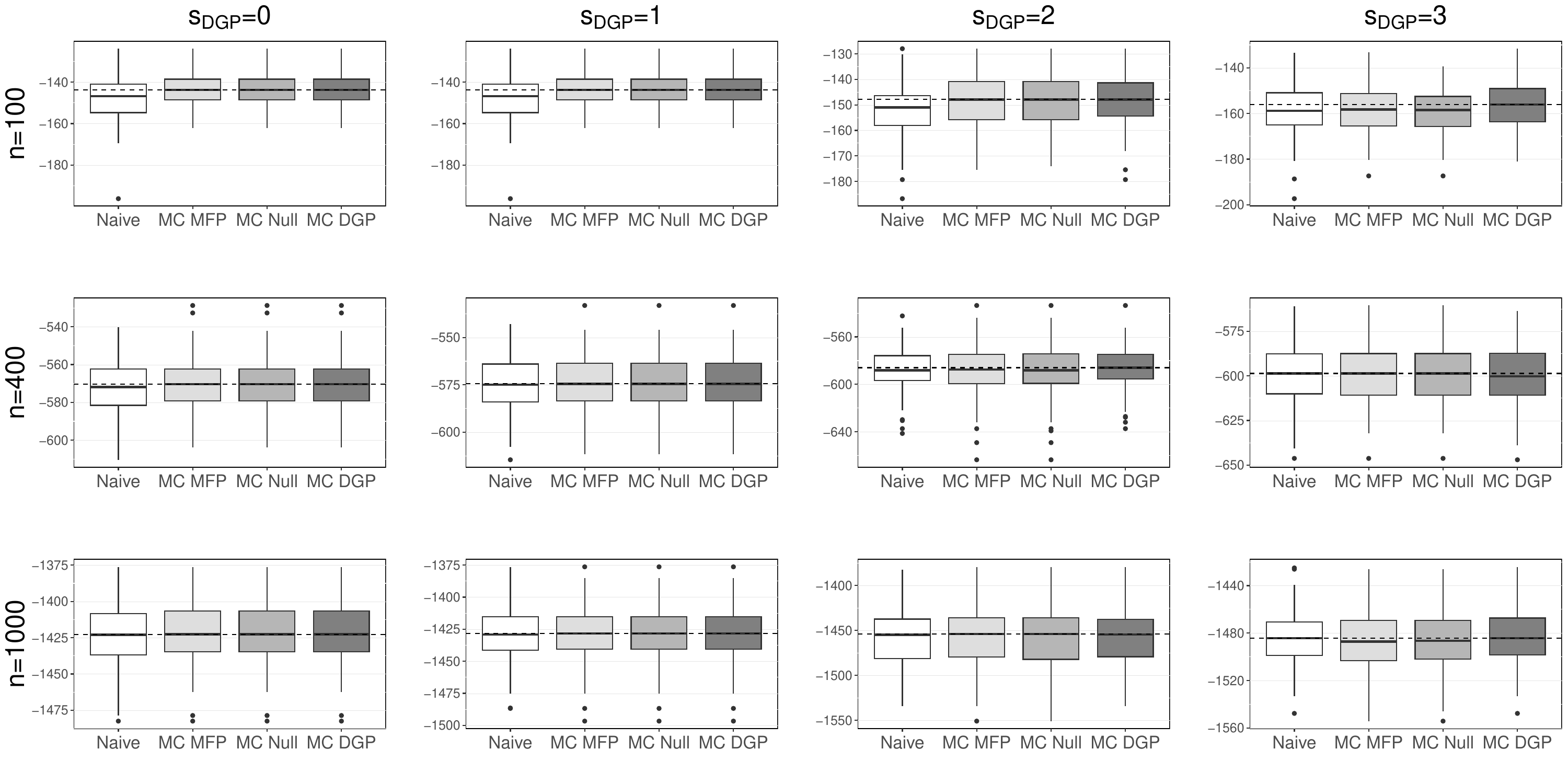}
\caption{Results of the simulation study: Predictive performance (scenario~1). The figure predictive log-likelihood values of TSVC models including $p=2$ covariates  with different methods for determining the DoF (Naive, MC MFP, MC Null, and MC DGP) for the BIC-based post-pruning fitted to samples of size $n\in\{100, 400, 1000\}$, where the true number of splits in the DGP was $s_{\text{DGP}}\in \{0, \dots, 3\}$.}
\label{ppl_p2}
\end{figure}

It is seen from Figure \ref{ppl_p2} that the large number of splits performed with the naive DoF approach came with a substantially decreased predictive performance compared to the other approaches in settings with low sample size ($n=100$). In these settings, the three MC-based approaches were superior to the naive approach and performed very similarly to each other. Note, however, that for the DGP with $s_{\text{DGP}} = 3$, where one of the splits was more difficult to detect due to decreased difference between the node-specific coefficients, MC DGP performed slightly better than the other MC-based approaches. For larger sample sizes the difference in predictive performance between the naive and the MC-based approaches vanished.

\subsection{Simulation with six covariates}

In the second scenario, data were generated based on the process
\begin{equation*}
Y = \beta_1(X_2)X_1 + \beta_2X_2 + \beta_3(X_4)X_3 + \beta_4 X_4 + \beta_5(X_6) + \beta_6X_6 + \varepsilon, 
\end{equation*}
where the error term $\varepsilon$ again followed a standard normal distribution. The (varying) coefficient of $X_1$ was given by $\beta_{1}(X_2)=0$ if $s_{\text{DGP}} = 0$ and $\beta_{1}(X_2)=I(X_2>0)$ if $s_{\text{DGP}} \geq 1$.  The coefficient of $X_3$ was given by $\beta_{3}(X_4) = 0$ if $s_{\text{DGP}} < 2$ and $\beta_{3}(X_4) = I(X_4>0)$ if $s_{\text{DGP}} \geq 2$. The coefficient of $X_5$ was defined as $\beta_5(X_6) = 0$ if $s_{\text{DGP}} <3$ and $\beta_{5}(X_6) = I(X_6>0)$ if $s_{\text{DGP}} = 3$. The non-varying linear coefficients were set to $\beta_{2} = \beta_{4} = \beta_{6} = 0$.

\begin{table}[ht]
\centering
\caption{Results of the simulation study: Number of splits (scenario 2). Average number of splits (with standard deviations in brackets) of TSVC models including $p=6$ covariates with different methods for determining the DoF (Naive, MC MFP, MC Null, and MC DGP) for the BIC-based post-pruning fitted to samples of size $n\in\{100, 400, 1000\}$, where the true number of splits in the DGP was $s_{\text{DGP}}\in \{0, \dots, 3\}$.}
\label{splits_p6}
\begin{tabularx}{\textwidth}{llYYYY}
\hline
$n$ & $s_{\text{DGP}}$ & \multicolumn{4}{c}{DoF approach}\\ 
& & Naive & MC MFP & MC Null & MC DGP \\
\hline
100  & 0 & 4.98 (0.20) & 0.00 (0.00) & 0.00 (0.00) & 0.00 (0.00) \\
     & 1 & 4.98 (0.14) & 0.05 (0.22) & 0.01 (0.10) & 0.95 (0.22) \\
     & 2 & 4.99 (0.10) & 0.06 (0.28) & 0.03 (0.14) & 1.19 (0.95) \\
     & 3 & 5.00 (0.00) & 0.03 (0.17) & 0.02 (0.14) & 0.77 (0.91) \\
\hline
400  & 0 & 4.55 (1.07) & 0.00 (0.00) & 0.00 (0.00) & 0.00 (0.00) \\
     & 1 & 4.79 (0.59) & 0.99 (0.10) & 0.96 (0.20) & 1.00 (0.00) \\
     & 2 & 4.93 (0.29) & 2.00 (0.00) & 1.97 (0.22) & 2.00 (0.00) \\
     & 3 & 5.00 (0.00) & 2.15 (0.36) & 2.02 (0.32) & 2.55 (0.50) \\
\hline
1000 & 0 & 3.66 (1.39) & 0.00 (0.00) & 0.00 (0.00) & 0.00 (0.00) \\
     & 1 & 4.32 (1.07) & 1.00 (0.00) & 1.00 (0.00) & 1.00 (0.00) \\
     & 2 & 4.75 (0.66) & 2.00 (0.00) & 2.00 (0.00) & 2.00 (0.00) \\
     & 3 & 4.89 (0.37) & 2.84 (0.37) & 2.69 (0.46) & 3.00 (0.00) \\
\hline
\end{tabularx}
\end{table}

Table \ref{splits_p6} shows that the naive approach yielded TSVC models with a severely inflated number of splits approaching the maximal number of splits ($S_{\text{max}} = 5$) even in settings where no splits where present. The strongly exaggerated number of splits resulting from the naive approach in this scenario confirms, again, that the search cost substantially increase with the number of covariates, in particular when compared to the previous scenario with $p=2$ covariates, where the number of splits was only moderately inflated. Analogously to the previous scenario, the number of splits with the naive approach decreased with sample size. In contrast, the MC-based approaches (MC MFP, MC Null, and MC DGP) overall underestimated the true number of splits, but very well approximated the true number of splits with increasing sample size. Specifically, MC MFP and MC Null resulted in almost no splits for $n=100$, whereas the MC DGP approach found approximately one split on average in all settings with $n=100$ and $s_{\text{DGP}}> 0$. It is worth noting that the average number of performed splits decreased in the setting with $s_{\text{DGP}} = 3$ compared to the setting with $s_{\text{DGP}}=2$ for all MC-based approaches. This behavior was likely caused by the increase in the variance of the response variable by adding a split to the DGP that makes splits harder to identify correctly by the TSVC algorithm. All MC-based approaches were close to the true number of splits for large sample size ($n=1000$) and a lower number of splits ($s_{\text{DGP}}\leq 2$) across all replications. For $s_{\text{DGP}}=3$, only the MC DGP approach selected the correct number of splits in all replications, whereas the MC MFP and the MC Null approaches sometimes selected TSVC models with a lower number of splits. Notably, the third split in the effect of $X_5$ was harder to detect due to the decreased difference of the linear coefficients between the leaf nodes compared to the previous two splits.  

\begin{figure}[!t]
\centering
\includegraphics[width = \textwidth]{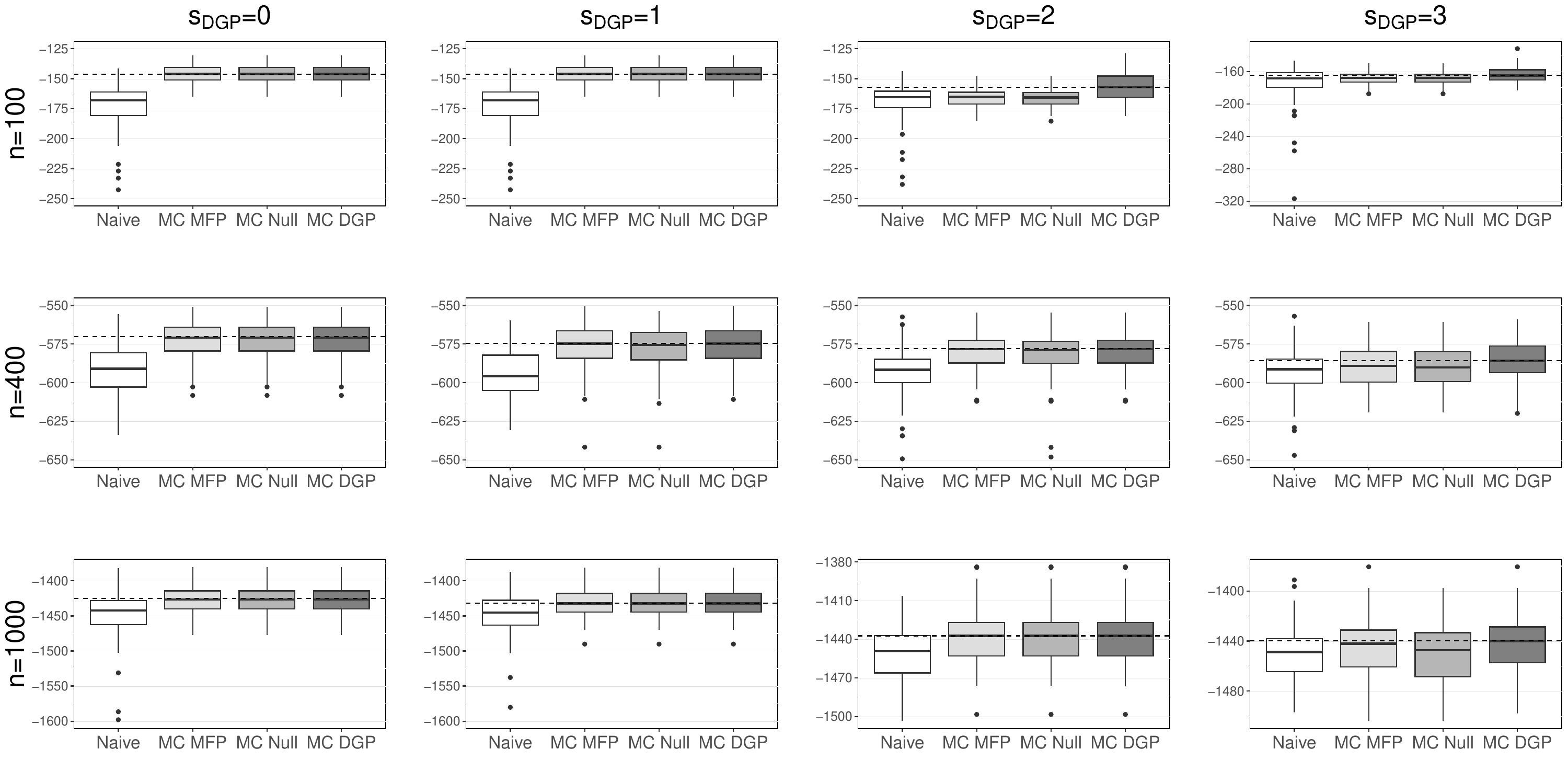}
\caption{Results of the simulation study: Predictive performance (scenario 2). The figure predictive log-likelihood values of TSVC models including $p=6$ covariates  with different methods for determining the DoF (Naive, MC MFP, MC Null, and MC DGP) for the BIC-based post-pruning fitted to samples of size $n\in\{100, 400, 1000\}$, where the true number of splits in the DGP was $s_{\text{DGP}}\in \{0, \dots, 3\}$.}
\label{ppl_p6}
\end{figure}

Figure \ref{ppl_p6} shows that the naive DoF approach resulted in strongly decreased predcitive log-likelihood values. The MC-based approaches performed substantially better and similarly to each other in most settings. However, the performance of the MC MFP and the MC Null approaches sightly deteriorated with increasing number of splits in the DGP compared to MC DGP. 

\subsection{Simulation with ten covariates}

In scenario 3, a DGP analogously to the previous scenario was applied. The only difference was that an additional set of four variables with linear effects $\beta_{7} = \beta_{8}=\beta_{9} = \beta_{10}=0$ were included as covariates and potential effect modifiers in the model fitting. 

\begin{table}[ht]
\centering
\caption{Results of the simulation study: Number of splits (scenario 3). Average number of splits (with standard deviations in brackets) of TSVC models including $p=10$ covariates with different methods for determining the DoF (Naive, MC MFP, MC Null, and MC DGP) for the BIC-based post-pruning fitted to samples of size $n\in\{100, 400, 1000\}$, where the true number of splits in the DGP was $s_{\text{DGP}}\in \{0, \dots, 3\}$.}
\label{splits_p10}
\begin{tabularx}{\textwidth}{llYYYY}
\hline
$n$ & $s_{\text{DGP}}$ & \multicolumn{4}{c}{DoF approach}\\ 
& & Naive & MC MFP & MC Null & MC DGP \\
\hline
100  & 0 & 5.00 (0.00) & 0.00 (0.00) & 0.00 (0.00) & 0.00 (0.00) \\
 & 1 & 5.00 (0.00) & 0.00 (0.00) & 0.00 (0.00) & 0.49 (0.50) \\
  & 2 & 5.00 (0.00) & 0.00 (0.00) & 0.00 (0.00) & 0.88 (0.46) \\
  & 3 & 5.00 (0.00) & 0.00 (0.00) & 0.00 (0.00) & 0.36 (0.69) \\
\hline
400  & 0 & 5.00 (0.00) & 0.00 (0.00) & 0.00 (0.00) & 0.00 (0.00) \\
  & 1 & 5.00 (0.00) & 0.76 (0.43) & 0.81 (0.39) & 1.00 (0.00) \\
  & 2 & 5.00 (0.00) & 1.86 (0.47) & 1.75 (0.58) & 2.00 (0.00) \\
  & 3 & 5.00 (0.00) & 1.69 (0.71) & 1.61 (0.69) & 2.25 (0.44) \\
\hline
1000 & 0 & 4.92 (0.42) & 0.00 (0.00) & 0.00 (0.00) & 0.00 (0.00) \\
 & 1 & 4.98 (0.14) & 1.00 (0.00) & 1.00 (0.00) & 1.00 (0.00) \\
 & 2 & 4.99 (0.10) & 2.00 (0.00) & 2.00 (0.00) & 2.00 (0.00) \\
 & 3 & 5.00 (0.00) & 2.29 (0.46) & 2.35 (0.48) & 3.00 (0.00) \\
\hline
\end{tabularx}
\end{table}

The results in Table \ref{splits_p10} again demonstrate that the naive DoF are not applicable (selecting the maximal number of splits $S_{\text{max}} = 5$ in almost all replications across all settings), because it neglects the considerable search cost that come with the increased number of $p=10$ covariates. The MC MFP and MC Null approach, on the other hand, overall underestimated the number of splits and even selected models with no splits in all the settings with low sample size ($n=100$). However, both approaches (MC MPF and MC Null) always correctly identified the true number of splits for $n =1000$ and $s_{\text{DGP}} \leq 2$, mirroring the results from the previous scenario. Also similar to scenario 2, the MC DGP approach resulted in a lower number of splits in case of lower sample sizes but always selected the correct number of splits for $n = 1000$. 

\begin{figure}[!t]
\centering
\includegraphics[width = \textwidth]{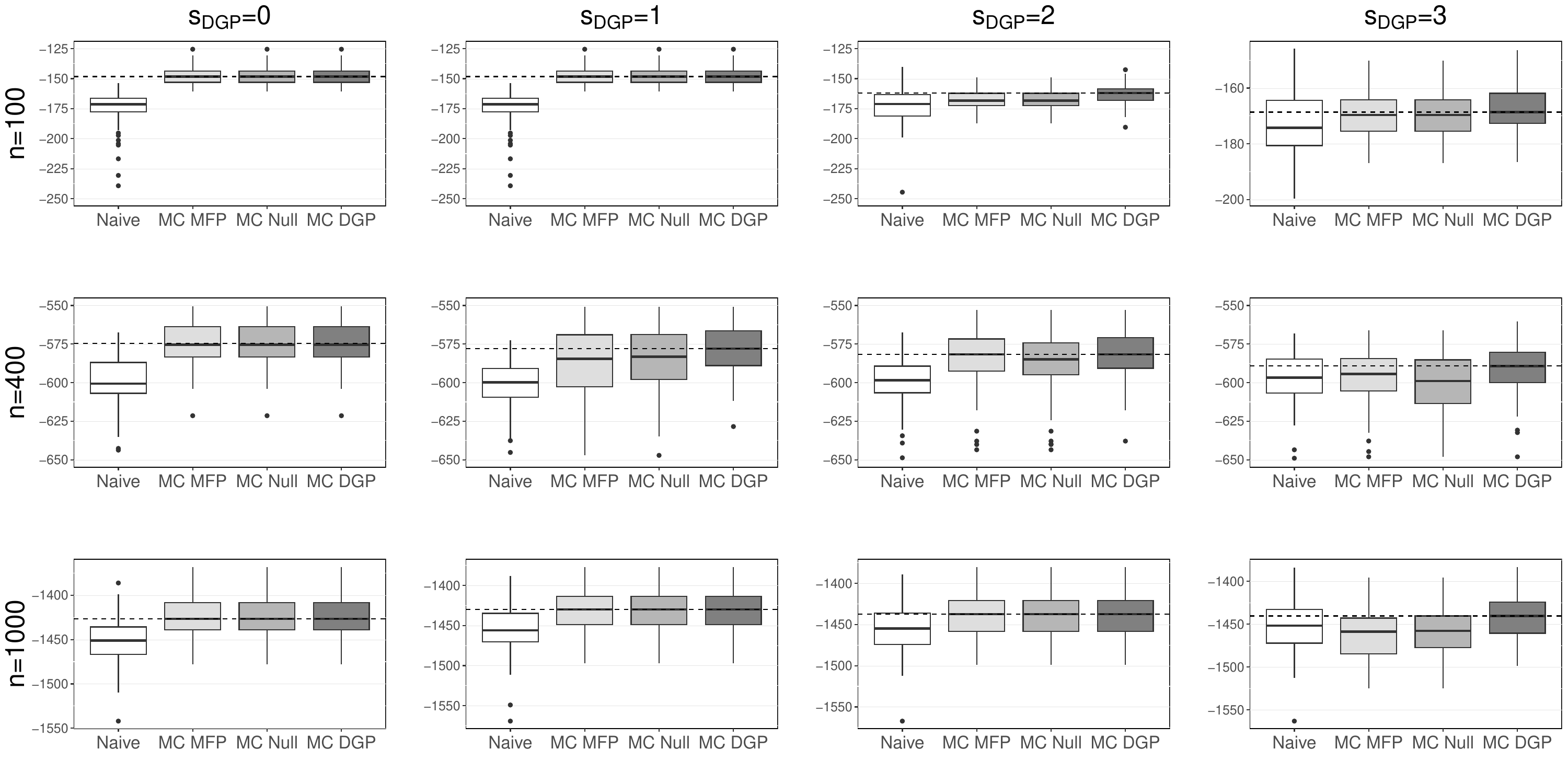}
\caption{Results of the simulation study: Predictive performance (scenario 3). The figure predictive log-likelihood values of TSVC models including $p=10$ covariates  with different methods for determining the DoF (Naive, MC MFP, MC Null, and MC DGP) for the BIC-based post-pruning fitted to samples of size $n\in\{100, 400, 1000\}$, where the true number of splits in the DGP was $s_{\text{DGP}}\in \{0, \dots, 3\}$.}
\label{ppl_p10}
\end{figure}

Figure \ref{ppl_p10} indicates that the naive DoF approach resulted in models with decreased predictive performance. The MC-based methods exhibited superior performance and yielded results that were generally similar to each other for $s_{\text{DGP}}\leq2$ with the MC DGP approach performing best. For the setting with $s_{\text{DGP}} = 3$ splits and sample size $n=1000$ in the DGP, where evidence for the third split in the data was lower, the MC MFP and the MC Null approaches exhibited worse performance than the MC DGP. The comparable performance of the naive approach demonstrates that overfitting in this setting does not cause to much harm in terms of prediction.

\subsection{Simulation based on the application}

In a final scenario of the simulation study, we considered data generated based on our application to the SHARE data. In particular, we considered data with $p=4$ standard normally distributed covariates and a sample of size $n=2985$ and fitted TSVC models with a maximal number of $S_{\text{max}} = 10$ splits. Values of the response variables were then generated from the process
\begin{equation*}
Y = \beta_1(X_2)X_1 + \beta_2X_2 + \beta_{3}(X_4) X_3 + \beta_{4}X_4 + \varepsilon\, ,
\end{equation*}
where $\varepsilon$ followed a standard normal distribution. The coefficient of $X_1$ was defined as $\beta_1(X_2) = 0$ if $s_{\text{DGP}} = 0$, $\beta_1(X_2) = I(X_2>0)$ if $s_{\text{DGP}} = 2$, $\beta_1(X_2) = I(X_2>0) + 2I(X_2>0.675)$ if $s_{\text{DGP}} = 4$, and $\beta_1(X_2) = I(X_2>0) +2I(X_2>0.675) - I(X_2<0.675)$ if $s_{\text{DGP}} = 6$. The effect of $X_3$ was defined analogously employing $X_4$ as the effect modifier. The linear main effects of $X_2$ and $X_4$ were set to $\beta_{2} = \beta_{4} = 0$. Note that we performed additional MC simulations to derive the DoF Null, as this data scenario is not represented in Table~\ref{edf}.

\begin{table}[ht]
\centering
\caption{Results of the simulation study: Number of splits (application-based data generation). Average number of splits (with standard deviations in brackets) of TSVC models including $p=4$ covariates with different methods for determining the DoF (Naive, MC MFP, MC Null, and MC DGP) for the BIC-based post-pruning fitted to samples of size $n=2\ 985$, where the true number of splits in the DGP was $s_{\text{DGP}}\in \{0, 2, 4, 6\}$.}
\label{splits_SHARE}
\begin{tabularx}{\textwidth}{lYYYY}
\hline
 $s_{\text{DGP}}$ & \multicolumn{4}{c}{DoF approach}\\ 
& Naive & MC MFP & MC Null & MC DGP \\
\hline
 0 & 1.01 (1.56) & 0.00 (0.00) & 0.00 (0.00) & 0.00 (0.00) \\
2 & 5.70 (2.56) & 2.13 (0.34) & 2.14 (0.35) & 2.13 (0.34) \\
4 & 8.00 (2.03) & 4.11 (0.31) & 4.12 (0.33) & 4.32 (0.47) \\
6 & 9.39 (1.00) & 6.44 (0.57) & 6.48 (0.61) & 7.04 (0.20) \\
\hline
\end{tabularx}
\end{table}

Similarly to the previous scenarios, the average number of splits reported in Table~\ref{splits_SHARE} indicate that the naive DoF approach resulted in an inflated number of splits, whereas the MC-based approaches yielded TSVC models with a number of splits much closer to the true number. All MC-based approaches slightly overestimated the true number of splits, similar to the results from scenario 1 with $p=2$, which suggests that these approaches tend to select a larger number of splits when only a few covariates are present. Interestingly, the approach applying the true DGP leads to the largest number of splits on average from among the MC-based approaches, in particular for $s_{\text{DGP}} = 6$, which indicates that an increased depth of the true trees makes tree fitting and selecting the correct split points more difficult, aggravating the disadvantage of the greedy algorithm.

\begin{figure}[!t]
\centering
\includegraphics[width = \textwidth]{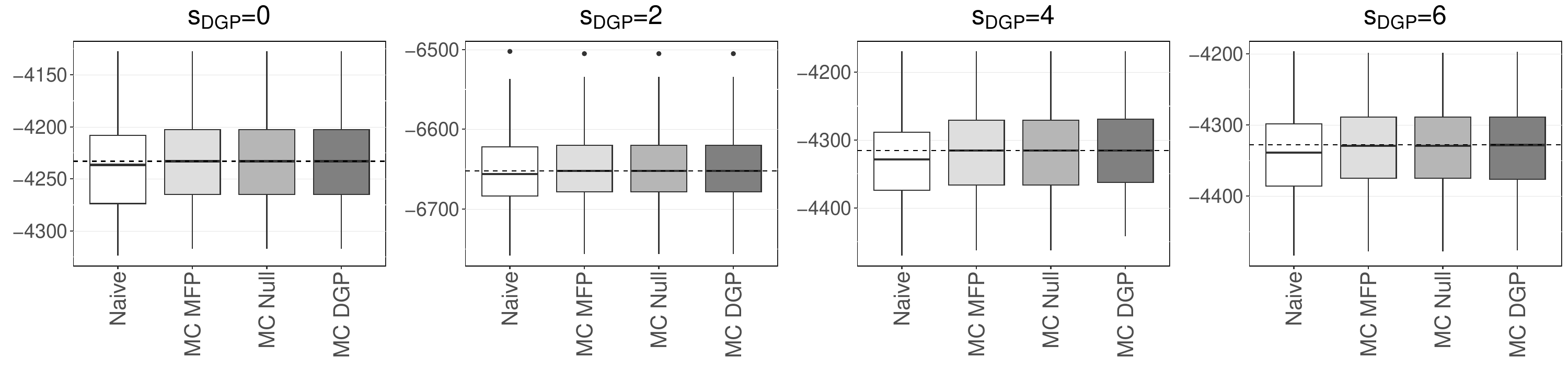}
\caption{Results of the simulation study: Predictive performance (application-based data generation). The figure predictive log-likelihood values of TSVC models including $p=4$ covariates  with different methods for determining the DoF (Naive, MC MFP, MC Null, and MC DGP) for the BIC-based post-pruning fitted to samples of size $n=2985$, where the true number of splits in the DGP was $s_{\text{DGP}}\in \{0, 2, 4, 6\}$.}
\label{ppl_SHARE}
\end{figure}

Figure \ref{ppl_SHARE} shows that the predictive performance of the models fitted using the MC-based DoF approaches was highly similar across all settings and superior to the naive approach, which deteriorated with a growing number of true splits. 

Overall, our results suggest that the BIC-based post-pruning using the proposed MC MFP approach to determine the DoF poses a reliable tool for TSVC model selection, in particular, given that the MC DGP approach can not be applied in practice. This is also the case for data outside the range of sample sizes and numbers of splits originally considered in our MC simulations for deriving the proposed formula in Equation~\eqref{mfpdf}.   

\section{Application}

For illustration, we applied a TSVC model with the BIC-based post-pruning strategy to data from SHARE. SHARE is a longitudinal, cross-national survey that collects data from individuals aged 50 years and older living in the European Union and Israel \citep{BoerschSupan2013}. Data collection for the first wave of SHARE started in 2004 in 19 different countries. Since then a total of nine waves have been conducted. The survey was mainly designed to provide information on how socio-economic and health-related factors influence the aging process. Here, we analyze data from the ninth wave collected from October 2021 to October 2022 in Germany \citep{Bergmann2024, Share2024}. The objective of our analysis was to investigate how quality of life (QoL) measured on an adapted 12-item version of the control, autonomy, self-realization and pleasure (CASP; \citealp{Hyde2003, BorratBesson2015}) scale is influenced by  socio-economic and health-related factors and to detect potential interaction effects between these factors. 

In a preliminary step of the analysis, for households with more than one individual participating in the survey, one representative was selected at random. This resulted in an analysis data set of  $n = 2,984$ individuals. The individual-level factors considered for modeling were: gender (male or female), age (in years), number of chronic diseases, and yearly household income (in Euro). In order to evaluate predictive performance of the fitted models on external observations, the ninth-wave SHARE data from Austria 
were considered. Summary statistics of these factors for the two countries are given in Table~\ref{tab:summary}. For more details on the ninth wave of SHARE, see also \citet{Bergmann2024} and \citet{Share2024}.

\begin{table*}[!t]
\caption{Analysis of the SHARE data: Summary. Summary statistics of the individual-level factors included in the analysis for Germany ($n=2\ 984$) and Austria ($n=2\ 360$)}\label{tab:summary}
\begin{center}
\begin{scriptsize}
\begin{tabularx}{\textwidth}{l l c *{6}{Y}}
\toprule
\bf{Variable}& \bf{Country} & \multicolumn{6}{c}{\bf{Summary statistics}}\\
\midrule
& & $x_{min}$ & $x_{0.25}$ & $x_{med}$ & $\overline{x}$ &$x_{0.75}$ & $x_{max}$ \\
Age (in years) &  Germany & 51 & 60  & 68  & 68.59 & 76 & \ \ 97 \\
 & Austria & 51 & 61 & 70 & 70.13 & 79 & 102 \\
Number of  & Germany & \  0 & \  1 & \  2 &  \ 2.16 &  \ 3 & \ \  12 \\
chronic diseases & Austria & \  0 & \  1 & \ 1 &  \ 1.85 & \ 3 & \ \  13 \\
Income (in Euro) & Germany & \  0 & 21, 400 & \ \ 34, 000 & \ \ 40, 869 & 51, 400 &  \ 457, 140 \\
& Austria & \  0  & 20, 125 & \ \ 30, 016 & \ \ 35, 322 &  44, 800 & \ 780, 200 \\
\end{tabularx}
\begin{tabularx}{\textwidth}{l l Y r}
Gender (female) \, & Germany & $1\, 628\, (54.54\%) $ &\\
 & Austria &  $1\, 459 \, (61.82\%) $ & \\
\bottomrule
\end{tabularx}
\end{scriptsize}
\end{center}
\end{table*}

We fitted a TSVC model with BIC-based post-pruning and the DoF calculated from Equation~\eqref{mfpdf} to the data. Confidence intervals (CIs) were constructed using the parametric bootstrap approach by \citet{Spuck2025} with $B=500$ bootstrap samples. The maximal number of splits in the TSVC models was set to $S_{\text{max}} = 10$. 

\begin{table*}[!t]
\caption{Analysis of the SHARE data: Model fit. Summary of the TSVC model with BIC-based post-pruning based on the MC MFP calculated DoF and $S_{\text{max}} = 10$ fitted to the wave-nine SHARE data from Germany. CIs were calculated based on $B= 500$ bootstrap samples }
\label{tab:model}
\begin{center}
\begin{scriptsize}
\begin{tabularx}{\textwidth}{l l YY}
\toprule
\textbf{Variable} & \textbf{Partition} & \textbf{Coefficient estimate} & \textbf{95\% CI} \\
\midrule
Gender (female) & --- &\ 0.210 & [-0.181; 0.559] \\
Age (in years) & Income $\leq 19, 400$ Euro &\ 0.006 & [-0.014;\ 0.046] \\
& Income $>19, 400$ Euro &\ 0.038 & [\ 0.180;\ 0.056] \\
Number of chronic diseases & --- & -1.039 & [-1.144;-0.935] \\  
Income (in 10,000 Euro) & --- & \ 0.267 & [\ 0.199;\ 0.407] \\
\bottomrule
\end{tabularx}
\end{scriptsize}
\end{center}
\end{table*}

Table \ref{tab:model} shows the coefficient estimates and CIs of the TSVC model.
It is seen that the algorithm performed one split in the effect of age. Increasing age was associated with a significantly increased QoL (at error level $\alpha = 0.05$) if the yearly household income was higher than $19, 400$ Euro, but this effect vanished for individuals with an income of $19, 400$ Euro or lower. More specifically, the results indicate that the QoL of an individual living in a household with a yearly income of more than $19, 400$ Euro is expected to increase by approximately 1 point ($\hat{\beta}=0.038$) on the CASP scale compared to a person that is 25 years younger (assuming all other covariates remain constant). Importantly, this effect is adjusted for the number of chronic diseases the person suffered from, which tends to increase with age and was found to significantly affect QoL. The expected CASP score is lowered by approximately 1 point ($\hat{\beta}=-1.039$) per chronic disease. Gender showed no significant effect on QoL, though women exhibited a slightly higher QoL compared to men. In addition, an increase in yearly household income was shown to be associated with an increased CASP score (by $\hat{\beta}=0.267$ per $10, 000$ Euro). 

\begin{table*}
\caption{Analysis of the SHARE data: External validation. Predictive log-likelihood values and number of performed splits of a linear (non-varying coefficients) model, a TSVC model with BIC-based post-pruning using Equation \eqref{mfpdf} to calculate the DoF, and a TSVC model with BIC-based post-pruning using the naive DoF approach fitted to the wave-nine SHARE data from Germany evaluated on the external observations from Austria}
\label{tab:exVal}
\begin{center}
\begin{footnotesize}
\begin{tabularx}{\textwidth}{l l Y Y}
\toprule
\textbf{Model} & \textbf{DoF appproach} & \textbf{Number of splits} & \textbf{Predictive log-likelihood} \\
\midrule
Linear & --- & --- & -7238.85 \\
TSVC & MC MFP & 1 & -7218.28 \\
TSVC & Naive & 9 & -7244.18 \\
\bottomrule
\end{tabularx}
\end{footnotesize}
\end{center}
\end{table*}

For comparison, we fitted also a classical linear model without varying coefficients and a TSVC model with BIC-based post-pruning using the naive DoF. Table \ref{tab:exVal} shows that the TSVC model with the MC MFP-based DoF presented above, which resulted in $s=1$ split, was most generalizable to the SHARE data from Austria compared to the other competitors. The naive DoF approach, conversely, resulted in a TSVC model with $s=9$ splits, which exhibited an even worse predictive performance than the linear model with no varying coefficients. These results demonstrate that TSVC models enable data-driven detection and easily interpretable modeling of relevant interaction effects, but post-pruning using the naive DoF approach may often result in too complex models. Post-pruning based on the proposed DoF approach, on the other hand, can lead to more parsimonious TSVC models with improved predictive ability.     

\section{Discussion and summary}

In this article, we developed an easy-to-apply formula to derive the DoF of a TSVC model. It can be used to calculate the DoF directly for any given combination of number of covariates, number of splits and sample size. This is the key benefit for practitioners compared to approaches that rely solely on MC simulations. 

In our simulation study we employed this formula for model selection via BIC and show that the proposed approach yields very accurate selection results. Well approximating the DoF resulted in more parsimonious models with a reduced number of splits and better predictive ability overall compared to the naive DoF approach, which neglects the search cost and tends toward overfitting in most settings. The proposed DoF approach was shown to be particularly beneficial in higher-dimensional data settings with a large number of covariates, where the naive approach yielded much larger models with a decreased predictive ability and an aggravated interpretation of the trees. The application to the SHARE data from Germany illustrated that the proposed DoF approach resulted in an easily interpretable TSVC model that was more generalizable to the external data from Austria than the the naive DoF approach or a classical linear model without varying coefficients.

Importantly, the true DoF of a data-driven model building depend on the form of the DGP. Here, we followed \citet{Wang2024} and estimated the DoF based on a null model without covariate effects or varying coefficients. \citet{Ye1998} argued that a DGP with no covariate effects leads to the largest DoF, because if no true but only erroneous (likely smaller) effects based on random chance can be found in the data, the search cost increases. This is supported by the findings in our simulations when comparing teh MC Null to the MC DGP approach. Therefore, our approach is rather conservative and may underestimate the number of splits, especially if the differences between the node-specific coefficients are small. 

Model selection performed in this article was based on the BIC, because of its tendency toward sparse models and well-known consistency property \citep{Schwarz1978, Zhang2023}. 
Alternative information criteria, like the AIC or DIC, may also be used in combination with the proposed DoF method for TSVC model building. More details on information criteria for model selection were given by \citet{Konishi1996} and \citet{Zhang2023}.

While the formula to approximate the DoF proposed here was specifically designed for TSVC models, our two-step methodology using the MC algorithm by \citet{Wang2024} and MFP regression as introduced by \citet{Royston1994} would generally be applicable for any model fitting procedure. The applied methodology may be a promising strategy for a systematic investigation of the DoF for other complex semi- and nonparametric modeling approaches, where closed-from expressions to calculate the DoF are currently unavailable, and poses an interesting topic for future research.   

\section*{Acknowledgements}

\begin{sloppypar}
Support by the German Research Foundation (DFG; grant: BE 7543/1-2) is gratefully acknowledged. \\

This paper uses data from SHARE Wave 9  (DOI: 10.6103/SHARE.w9ca900) see \citet{BoerschSupan2013} for methodological details.
The SHARE data collection has been funded by the European Commission, DG RTD through FP5 (QLK6-CT-2001-00360), FP6 (SHARE-I3: RII-CT-2006-062193, COMPARE: CIT5-CT-2005-028857, SHARELIFE: CIT4-CT-2006-028812), FP7 (SHARE-PREP: GA N$^{\circ}$ 211909, SHARE-LEAP: GA N$^{\circ}$ 227822, SHARE M4: GA N$^{\circ}$ 261982, DASISH: GA N$^{\circ}$ 283646) and Horizon 2020 (SHARE-DEV3: GA N$^{\circ}$ 676536, SHARE-COHESION: GA N$^{\circ}$ 870628, SERISS: GA N$^{\circ}$ 654221, SSHOC: GA N$^{\circ}$ 823782, SHARE-COVID19: GA N$^{\circ}$ 101015924) and by DG Employment, Social Affairs \& Inclusion through VS 2015/0195, VS 2016/0135, VS 2018/0285, VS 2019/0332, VS 2020/0313, SHARE-EUCOV: GA N$^{\circ}$101052589 and EUCOVII: GA N$^{\circ}$101102412. Additional funding from the German Federal Ministry of Education and Research (01UW1301, 01UW1801, 01UW2202), the Max Planck Society for the Advancement of Science, the U.S. National Institute on Aging (U01\_AG09740-13S2, P01\_AG005842, P01\_AG08291, P30\_AG12815, R21\_AG025169, Y1-AG-4553-01, IAG\_BSR06-11, OGHA\_04-064, BSR12-04, R01\_AG052527-02, R01\_AG056329-02, R01\_AG063944, HHSN271201300071C, RAG052527A) and from various national funding sources is gratefully acknowledged (see www.share-eric.eu).
\end{sloppypar}

\section*{Statements and declarations}

\textbf{Competing interests:} The authors have no competing interests to declare.

\bibliographystyle{plainnat}
\bibliography{bib}

\newpage

\appendix

\section{Multivariable fractional polynomials}

In the following, we outline details on the the MFP fitting procedure~\citep{Royston1994, Sauerbrei1999}. As a preprocessing step covariates are ordered from most to least relevant. Specifically, $p$ models with linear effects for all covariates excluding each covariate once are fitted to the data and the order of the covariates is determined based on the values of the corresponding model test statistics (comparing each model to the full linear model).  Following this order and starting with the covariate $X_j$ with the largest test statistic, the procedure evaluates fractional polynomials with degree $d\leq 2$ of the form 
\begin{equation}
    f(X_j)=\sum_{k=1}^{d}\xi_{jk}X_j^{p_{jk}}\, ,
\end{equation}
where $p_{jk}\in \{-2, -1, -0.5, 0, 0.5, 1, 2, 3\}$ and $x_j^{p_{jk}} = \text{log}(X_j)$ if $p_{jk} = 0$ for $k = 1,\dots , d$. If $p_{j1} = p_{j2}$, the fractional polynomial is given by $f(X_j) = \xi_{j1}X_{j}^{p_{j1}} + \xi_{j2}X_{j}^{p_{j2}}\text{log}(X_j)$. In step 1, a model including the effect of $X_j$ using the best functional form from all possible fractional polynomials of degree $d=2$ (based on the minimal deviance) is tested against a model with a null effect for~$X_j$, while linear effects are assumed for all other covariates. If the test indicates no significantly improved performance (at a prespecified error level $\alpha$), $X_j$ is excluded and the procedure continues analogously with step 1 and the next covariate in the order. If not, in the second step, the model with the best fractional polynomial of degree $d=2$ is tested against a model with a linear effect of $X_j$. If the test yields no significant result, a linear effect of $X_j$ is included in the model and the algorithm starts from step 1 for the next variable in the order. Otherwise, in the third step, models with the best performing fractional polynomials of degrees $d=1$ and $d=2$ (based on the minimal deviance) are compared analogously to the previous steps. In case of significance, the fractional polynomial of degree $d=2$ is selected. Otherwise, the fractional polynomial with $d=1$ is selected as functional form for $X_j$. The process is subsequently repeated for the next covariate in the order, while the functional forms of previously selected covariates is kept. This procedure is iterated until the selected functional forms for each of the covariates remain the same between iterations.

\end{document}